\documentclass[aps,prl,preprint]{revtex4-2}
\usepackage{CJK}
\usepackage[utf8x]{inputenc}
\usepackage[T1]{fontenc}
\usepackage{mathptmx} % Use Times Font

\usepackage[pdftex]{graphicx} % Required for including pictures
\usepackage[english]{babel} % Swedish translations
\usepackage{calc} % To reset the counter in the document after title page
\usepackage{enumitem} % Includes lists

\usepackage[protrusion=true,expansion=true]{microtype} % Improves typography, load after fontpackage is selected

\usepackage{lipsum} % Used for inserting dummy 'Lorem ipsum' text into the template
\usepackage{siunitx}
\usepackage{multirow}
\usepackage{amsmath}
\usepackage{caption}
\usepackage{subcaption}
\usepackage[raggedright]{sidecap}
\usepackage{nicematrix}

\begin{document}
\begin{CJK*}{GB}{} % Use default fonts from CJK (see below)
\title{Enhancement of the photon pair generation rate by spontaneous four-wave mixing in bent waveguides}
\author{Maria Paszkiewicz}
\affiliation{Karlsruhe Institute of Technology, Germany}
\author{Carsten Rockstuhl}
\affiliation{Karlsruhe Institute of Technology, Germany}
\date{\today}

\begin{abstract}
Enhancing nonlinear optical effects is critical to improving the performance of many functional devices for nonlinear and quantum optical applications. Here we study the possibility of bending a waveguide to enhance the photon pair generation rate in a spontaneous four-wave mixing process. Whereas intuition might suggest that bending makes the process less efficient, because the waveguides are more lossy when bent, we show that the increase in the effective nonlinearity outperforms the disadvantage for moderate bending radii. That can be explained by a better localization of the guided eigenmodes, leading to a reduced mode area. By studying selected waveguide devices with a fixed length, we demonstrate an optimal improvement between 5~\% and 270~\% in the photon pair generation rate for an optimal bending radius. These findings have implications for the future design of integrated photonic devices for quantum optical applications, especially in cases where the chip estate tends to be a limiting factor.
\end{abstract}

\maketitle

\section{Introduction}

Photon pair generation in waveguides or ring resonators leads to
the creation of squeezed states of light. This capability can enable
on-chip quantum photonic technologies \cite{borghi2017nonlinear, azzini2012ultra}, because the photon pair sources can be implemented in compact and complex quantum circuits \cite{politi2008silica, davanco2012telecommunications}. It has the advantage that the photon pairs do not need to be injected from an external source, and so coupling loss can be avoided \cite{shin2023photon}. The on-chip creation of entangled photon pairs, demonstrated for example in \cite{silverstone2014chip} with two \SI{5.2}{mm} long spiral waveguides or two microring resonators \cite{silverstone2015qubit}, can be used for quantum interferometry. This paves the way to quantum communications and optical quantum computation \cite{politi2009integrated, chi2022programmable}, e.g. Gaussian boson sampling \cite{zhong2020quantum}, as well as quantum metrology to light-sensitive samples \cite{crespi2012measuring}. It has been theoretically predicted and experimentally verified that spontaneous four-wave mixing (SFWM) in a third-order nonlinear material is an efficient source for quantum correlated photon pairs \cite{shin2023photon, alibart2006photon}. Third-order nonlinear processes in silicon like four-wave mixing (FWM), self- and cross-phase modulation (SPM, XPM) or two-photon absorption (TPA) can also be exploited for ultrafast all-optical switching at low power \cite{koos2007nonlinear}, temporal imaging \cite{salem2008optical}, optical couplers, and logic gates \cite{fujisawa2006all}.\\

Generally speaking, enhancing the nonlinearities can increase the efficiency of all these signal-processing devices in integrated photonic circuits. Since the nonlinearity of the material is an intrinsic quantity and hard to change, the most obvious solution is to make the nonlinear waveguides longer. However, that is not always desirable since chip estate is precious. Especially with an eye on an ever-denser integration of photonic components on chips, it would be desirable to maintain the spatial footprint of the nonlinear components small but enhance the nonlinear response. In that respect, a systematic bending of the waveguides has already been suggested, and it was shown that the waveguide length can be increased up to several centimeters by forming low-loss spirals using smooth Euler bends \cite{moss2013new, cherchi2014low}. These specifically bent waveguides minimize higher-order mode coupling \cite{morrison2016four}.\\  

In different integrated photonic platforms (with varying core sizes of the waveguides and considered refractive indices), waveguides can be bent with a reasonable radius without significantly increasing losses \cite{cherchi2014low, ebers2018oblique}. This property was already exploited in literature \cite{cherchi2014low, morrison2016four, moss2013new}. However, so far, the possible bend has only been considered from the perspective of increasing the waveguide length without compromising the spatial footprint, leaving the impact of the bend on the waveguide nonlinearity completely unattended. An effect is safe to be expected, as the modal properties change when the waveguide is bent, and most notably, the field localization improves. This translates directly to the effective nonlinearity of the waveguide, which promises to render any nonlinear effect more efficient. However, the fundamental question remains whether these benefits outweigh the detrimental aspect of increased losses when the waveguides are bent. This contribution strives to answer this question while concentrating on the important example of the photon pair generation rate in a spontaneous four-wave mixing process.\\ 

To give a comprehensive answer, we consider multiple waveguide systems that could be considered for this purpose and rely on different material platforms. The manuscript is structured into various sections to explore the problem comprehensively. In section \textit{Theory}, we describe at first the formalism used to calculate the nonlinear coefficients and photon pairs generation rate for channel waveguides. This serves as a reminder, and the general framework has been presented in the literature \cite{review_article, milica_article_draft}. In dedicated sections, we discuss the influence of different types of losses, outline the different waveguide systems we considered, and describe the numerical details of our procedure. Finally, in the \textit{Results} section, we present the results of our calculations for commonly used nonlinear waveguides with different curvatures, and we perform a study on a lossy waveguide to show the peak in nonlinear response. We present the results in detail for two different material systems and provide the results for several waveguide systems in an aggregated table. We finish with the physical discussion of the achieved results.

\section{Theory}\label{theory}
Four-wave mixing is a typical third-order ($\chi_3$) nonlinear optical process, where two or three wavelengths interact within a medium to generate a fourth wavelength. Four-wave mixing can occur classically or spontaneously, depending on whether an external input signal is provided or not in addition to a pump \cite{moss2013new}. In spontaneous four-wave mixing, the case on which we concentrate here, the annihilation of two pump photons results in the creation of a pair of one signal and one idler photon. The created photons are usually entangled, and the superposition of created pairs of photons forms squeezed light states. These two effects are of interest in quantum information processing protocols \cite{review_article}.\\

The generation rate $R$ of photon pairs can be expressed as 
\begin{equation}\label{eq:generation_rate}
    R = \frac{\text{number of generated pairs}}{\mathrm{time}} = \frac{2\pi}{\hbar^2}\int \int \delta(\Omega(\omega_\text{S},\omega_\text{I})) |M(\omega_\text{S},\omega_\text{I})|^2 \frac{1}{v_\text{S}}\frac{1}{v_\text{I}}\mathrm d \omega_\text{S} \mathrm d \omega_\text{I}\,,
\end{equation}
which is analogous to the formula (A16) in \cite{milica_article_draft}, when integrating over the frequencies instead of the wavenumbers. In the equation above, $\hbar$ is the reduced Planck constant, $\omega_\text{S}$ and $\omega_\text{I}$ are the angular frequencies of signal and idler photons, respectively, and $v_\text{S}$ and $v_\text{I}$ refer to the group velocity of signal and idler mode, respectively. The quantity $M(\omega_\text{S},\omega_\text{I})$ with units m$^3$s$^{-2}$kg corresponds to the FWM generation efficiency \cite{borghi2017nonlinear}. The Dirac delta distribution $\delta(x)$ depends on $\Omega(\omega_\text{S}, \omega_\text{I})$, which expresses the frequency difference of the annihilated and created photons $\Omega(\omega_\text{S}, \omega_\text{I})=2\omega_\text{P} - \omega_\text{S} - \omega_I$, for a fixed frequency of the pump photons $\omega_\text{P}$. We use the condition $\Omega(\omega_\text{S}, \omega_\text{I})=0$ to express energy conservation. We require the unit of the photon pair generation rate to be [Hz], i.e., pairs/s. This formula was derived in \cite{review_article} and \cite{milica_article_draft} from Fermi's golden rule with continuous-wave (CW) pump, after defining interaction Hamiltonian for creating signal and idler photons, and annihilating pump photons.\\

The crucial quantity to calculate is the FWM
generation efficiency $M(\omega_\text{S}, \omega_\text{I})$. It was shown in \cite{milica_article_draft} that it can be expressed as 
\begin{equation}\label{eq:M_first}
M(\omega_\text{S}, \omega_\text{I}) = \frac{6\pi P_\text{P}}{\varepsilon_0 \hbar \omega_\text{P} v_\text{P}}K(\omega_\text{S},\omega_\text{I},\omega_\text{P},\omega_\text{P})\,,
\end{equation}
where $P_\text{P}$ and $v_\text{P}$ are the power and the group velocity of the pump source, respectively, and $\varepsilon_0$ is the vacuum permittivity. We can find in analogy to equation (11) in \cite{milica_article_draft} that $K(\omega_\text{S},\omega_\text{I},\omega_\text{P},\omega_\text{P})$ reads as 
\begin{equation}\label{eq:K}
K(\omega_\text{S},\omega_\text{I},\omega_\text{P},\omega_\text{P}) = \int \Gamma_3^{ijkl}(\mathbf r) (D_\text{S}^i(\mathbf r))^*(D_\text{I}^j(\mathbf r))^*D_\text{P}^k(\mathbf r)D_\text{P}^l(\mathbf r) \mathrm d \mathbf r\,,
\end{equation}
where
\begin{equation}\label{eq:Gamma}
\Gamma_3^{ijkl}(\mathbf r) = \frac{\chi_3^{ijkl}(\mathbf r)}{\varepsilon_0^2\varepsilon_1^2(\mathbf r;\omega_\text{P})\varepsilon_1(\mathbf r;\omega_\text{I})\varepsilon_1(\mathbf r;\omega_\text{S})}
\end{equation}
expresses a normalized nonlinearity, where the third-order nonlinear susceptibility tensor $\chi_3^{ijkl}(\mathbf r)$ is normalized to the electric permittivity of the waveguide $\varepsilon_1(\mathbf{r};\omega_\text{J})$. We consider here centrosymmetric materials, for which $\chi_2^{ijk}(\mathbf r)$ disappears. Superscript indicates Cartesian coordinates, which are summed over when repeated. \\

The electric displacement field for a waveguide mode that we have to consider in the expression \eqref{eq:K} is, following Eqn. (4) from \cite{milica_article_draft}, defined as
\begin{equation}\label{eq:D}
\mathbf D_\mathrm{J}(\mathbf r) = \sqrt{\frac{\hbar\omega_\mathrm{J}}{4\pi}}\mathbf d_\mathrm{J}(\mathbf{r}_\perp)e^{\mathrm i \beta_\mathrm{J}z}\,.
\end{equation}
In this expression, the displacement field amplitude of the mode $\mathbf d_\mathrm{J}(r_\perp)$ is defined in a plane perpendicular to the principal propagation direction $z$ of the mode J. The propagation constant of the mode J is denoted by $\beta_\text{J}$. The subscript J denotes the pump P, signal S, or idler I, i.e., $\text{J}=\{\text{P},\text{S},\text{I}\}$. The displacement field amplitudes should be normalized according to equation (70) from \cite{review_article}
\begin{equation}\label{eq:normalization}
\int \frac{\mathbf d^*_\mathrm{J}(x,y)\cdot\mathbf d_\mathrm{J}(x,y)}{\varepsilon_0\varepsilon_1(x,y;\omega_\mathrm{J})}\frac{v_\mathrm{ph}(x,y;\omega_\mathrm{J})}{v_\mathrm{g}(x,y;\omega_\mathrm{J})}\mathrm dx\mathrm dy = 1.
\end{equation}
The quantities $v_\mathrm{ph}(x,y;\omega_\mathrm{J})$ and $v_\mathrm{g}(x,y;\omega_\mathrm{J})$ denote the local phase velocity and group velocity, respectively. They are derived from the bulk properties of the material that is at a specific point in the waveguide and account for material dispersion \cite{banic2022resonant}.\\

Considering the formula for the displacement field \eqref{eq:D} and the expression \eqref{eq:Gamma}, we can rewrite $K(\omega_\text{S},\omega_\text{I},\omega_\text{P},\omega_\text{P})$ as
\begin{multline}\label{eq:K_rewritten}
K(\omega_\text{S},\omega_\text{I},\omega_\text{P},\omega_\text{P}) = \frac{\hbar^2\sqrt{\omega_\text{S}\omega_\text{I}}\omega_\text{P}}{16\pi^2}\int_{-\frac{L}{2}}^{\frac{L}{2}} e^{\mathrm i z(2\beta_\text{P}-\beta_\text{S}-\beta_\text{I})}\mathrm dz \\
\cdot \varepsilon_0^2\int_\mathcal{D} \chi_3^{ijkl}(x,y) \left( e_\text{S}^i(x,y)e_\text{I}^j(x,y) \right)^*e_\text{P}^k(x,y)e_\text{P}^l(x,y)\mathrm dx\mathrm dy\,,
\end{multline}
where we consider the wave propagating over length $L$.
In the derivation of the above formula, we already took into account the relation between the electric and the displacement field: \mbox{$d_\text{J}^i(\mathbf r_\perp)=\varepsilon_0\varepsilon_1(\mathbf r_\perp; \omega_\text{J}) e_\text{J}^i(\mathbf r_\perp)$} and
\begin{multline}    
\int \Gamma_3^{ijkl}(\mathbf r_\perp) d_P^i(\mathbf r_\perp)d_P^j(r_\perp)d_\text{S}^{k*}(\mathbf r_\perp)d_\text{I}^{l*}(\mathbf r_\perp) \mathrm d \mathbf r_\perp \\
=\varepsilon_0^2\int \chi_3^{ijkl}(x,y) \left( e_\text{S}^i(x,y)e_\text{I}^j(x,y) \right)^*e_\text{P}^k(x,y)e_\text{P}^l(x,y) \mathrm dx\mathrm dy\,,
\end{multline} % It was divided by 4 but I don't see it now so I deleted it.
following from the considerations (380-382) in \cite{review_article}, appropriately for channel waveguides. Integrating over $\mathbf{r}_\perp= \mathrm dx \mathrm dy$ means integrating over the entire plane $\mathcal D$ perpendicular to the propagation direction of the waves.\\

To simplify the formula \eqref{eq:K_rewritten} further, we introduce the nonlinear coefficient \cite{review_article} for channel waveguides
\begin{equation}\label{eq:gamma_SFWM}
\gamma_{\text{SFWM}} = \frac{3(\omega_\text{S}\omega_\text{I}\omega_\text{P}^2)^{1/4}\varepsilon_0}{4\sqrt{v_\text{S}v_\text{I}v_\text{P}^2}} \int \chi_3^{ijkl}(x,y) \left( e_\text{S}^i(x,y)e_\text{I}^j(x,y) \right)^*e_\text{P}^k(x,y)e_\text{P}^l(x,y)\mathrm dx\mathrm dy\,.
\end{equation}
The nonlinear coefficient $\gamma_\mathrm{SFWM}$ is a measure of nonlinearity in a given structure corresponding to the strength of spontaneous four-wave mixing \cite{morrison2016four}. It helps to compare the nonlinearity of different structures and materials. The nonlinear coefficient is often defined with the use of an effective area of the waveguide mode \cite{review_article, leuthold2010nonlinear, koos2007nonlinear}

\begin{equation}\label{eq:Aeff}
A_\text{eff} = \frac{ N_\text{P}N_\text{P}N_\text{I} N_\text{S}}{\int \frac{\chi_3^{ijkl}(x,y)}{\bar{\chi}_3}\left[ e_\text{P}^i(x,y)e_\text{P}^j(x,y) \right]^*e_\text{I}^k(x,y)e_\text{S}^l(x,y)\mathrm dx\mathrm dy}\,,
\end{equation}
where
\begin{equation}\label{eq:N}
N_\text{J} = \sqrt{\int \frac{n(x,y;\omega_\text{J})/\bar{n}_\text{J}}{v_\mathrm{g}(x,y;\omega_\text{J})/v_\text{J}}\mathbf e^*_\text{J}(x,y)\cdot \mathbf e_\text{J}(x,y) \mathrm dx\mathrm dy}\,,
\end{equation}
with $\bar{n}_\text{J}$ being the value of the refractive index of the core at wavelength $\lambda_\text{J}$ and $\bar{\chi}_3$ -- the characteristic value of the susceptibility tensor components of the core. This derivation utilized the normalization condition \eqref{eq:normalization}. With the formula \eqref{eq:Aeff}, the nonlinear coefficient simplifies to

\begin{equation}\label{eq:gamma_simplified}
\gamma_\text{SFWM} = \frac{3(\omega_\text{S}\omega_\text{I}\omega_\text{P}^2)^{1/4}\bar{\chi}_3}{4\varepsilon_0\sqrt{\bar{n}_\text{S}\bar{n}_\text{I}\bar{n}_\text{P}\bar{n}_\text{P}}c^2}\frac{1}{A_\text{eff}}\,.
\end{equation}
    
Utilizing all of the above equations results in 
\begin{equation}\label{eq:K_final}
K(\omega_\text{S},\omega_\text{I},\omega_\text{P},\omega_\text{P}) = \frac{\hbar^2(\omega_\text{S}\omega_\text{I}\omega_\text{P}^2)^{1/4}\varepsilon_0}{12\pi^2}\sqrt{v_\text{S}v_\text{I}v_\text{P}^2}\; \gamma_\text{SFWM} \int_{-L/2}^{L/2}e^{\mathrm i z\Delta \beta}\mathrm dz\,,
\end{equation}
where the phase matching (momentum conservation) is quantified with \mbox{$\Delta \beta=2\beta_\text{P} - \beta_\text{S} - \beta_\text{I}$}. Finally, we arrive at the expression for $M(\omega_\text{S},\omega_\text{I})$
\begin{equation}\label{eq:M}
M(\omega_\text{S},\omega_\text{I}) = \frac{P_{\text P} \hbar(\omega_\text{S}\omega_\text{I})^{1/4}}{2\pi}\sqrt{\frac{v_\text{S}v_\text{I}}{\omega_\text{P}}}\gamma_\text{SFWM}L\mathrm{sinc}\left(\Delta \beta \frac{L}{2}\right)\,,
\end{equation}
where we used $\mathrm{sinc}(x) = \sin(x)/x$.
    
\subsection{Treatment of losses}
In the above derivation, neither scattering nor absorption losses were included explicitly. In this subsection, we discuss the various sources of losses and their significance for the considered waveguide materials and structures.\\

One common energy limitation is two-photon absorption (TPA) quantified by the coefficient \mbox{$\beta_\text{TPA}=\frac{3\omega}{2\varepsilon_0c^2n^2}\Im\{\chi^{(3)}\}$}, governed by the imaginary part of the nonlinear susceptibility \cite{osgood2009engineering}. The generated photon pairs could be lost due to TPA. Especially silicon-on-insulator (SOI) suffers from TPA \cite{liang2004nonlinear} if the photon frequencies (pump, signal, or idler) are higher than the silicon bandgap frequency \cite{borghi2017nonlinear}, which is the case at telecommunication wavelengths 
\cite{apiratikul2014nonlinear}. The solution can be to shift the pump to a longer wavelength or use another material like silicon nitride, which has a bigger band gap and is characterized by low nonlinear and linear losses at telecommunication wavelengths \cite{moss2013new}. TPA, free carrier absorption (FCA), and dispersion \cite{driscoll2011directionally} can also be reduced through lower pump powers. It was shown experimentally \cite{clemmen2009continuous} that the correlated photons can still be distinguished from accidental coincidences for a CW pump power of 10~mW at wavelengths close to \SI{1.55}{\micro\meter}. In \cite{driscoll2011directionally}, it was shown that the pump powers around 100~mW are still below conversion efficiency (the ratio between the output idler power and input probe power) enhancement saturation due to TPA and FCA at telecommunications wavelengths \cite{driscoll2011directionally}. In our computations, we use a pump power of 100~mW. However, due to the square proportionality of the photon pair generation rate to the pump power, our findings should apply to smaller powers as well.\\

Another issue are propagation losses caused by scattering on sidewalls due to surface roughness \cite{hammer2024estimation}. Such surface roughness can be caused, e.g., by an imperfect etching process in the waveguide fabrication \cite{morrison2016four, Sakai2001}. In silicon nanowires, this effect can worsen the generation of high absolute idler powers \cite{morrison2016four}. However, we do not consider fabrication imperfections in this theoretical study.\\

In this work, we consider only the propagation losses along waveguides induced by bents. The bending loss is included in the integral (\ref{eq:K_rewritten}) in the form $\int_{-L/2}^{L/2}e^{\mathrm i \Re\{\Delta \beta\}z}e^{-\beta_\text{im}z}\mathrm dz$, where the total imaginary part of the propagation constant equals $\beta_\text{im} = 2|\Im\{\beta_\text{P}\}| + |\Im\{\beta_\text{S}\}| + |\Im\{\beta_\text{I}\}|$. The absolute values are introduced to make sure the imaginary parts are positive regardless of the convention used. Integrating the exponential part of the amplitudes from $-L/2$ to $L/2$ results in
\begin{equation}\label{eq:loss_integral}
    \int_{-L/2}^{L/2}e^{\mathrm i \Delta \beta z}e^{-\beta_\text{im}(z+L/2)}\mathrm dz = \frac{2e^{-\frac{L}{2}\beta_\text{im}}\sinh\left( \frac{L}{2}(\beta_\text{im}-\mathrm i \Delta \beta)\right)}{\beta_\text{im}-\mathrm i \Delta \beta}\,.
\end{equation}
Note that the term in the expression above that expresses the losses, i.e., the second exponential on the left-hand side of the equation, takes a distance from 0 to $L$ as the effective argument. We ensure with this formulation that the losses are properly accumulated and consider the propagation in a waveguide section of length $L$. When the phase-matching condition is satisfied, the peak photon pair generation rate occurs for $\Delta \beta=0$. In this case, the integral \eqref{eq:loss_integral} approaches $-(\exp(-\beta_\text{im}L)-1)/\beta_\text{im}$, which is analogous to formula (30) in \cite{milica_article_draft}.\\

We are aware that in the experiments, the signal-to-noise ratio SNR = $(c_a+c_p)/c_a$, where $c_a$ is the number of accidental coincidences and $c_p$ is the number of coincidences due to correlated photons \cite{clemmen2009continuous}, may be affected by free carrier absorption and bending loss. The quantum information application of generated photon pairs may be limited as the heralding of photons may be inaccurate due to the loss of either the signal or idler photon \cite{davanco2012telecommunications, nunn2021heralding}. Fortunately, the linear loss (bending loss) is low (if not negligible) in the case of silicon \cite{Sakai2001} and silicon nitride. This is characteristic for structures with high index contrast \cite{Sakai2001}, where the guided mode is strongly confined to the waveguide core. It will be discussed in the \textit{Results} section.\\

The last aspect of practical interest is the photon pair heralding efficiency, which quantifies the probability that if an idler photon were detected, the signal photon would also be detected (and vice-versa). In the presence of propagation losses, this efficiency would be impaired because of the loss of either idler or signal photon from the pair. The influence of absorption loss on heralding efficiency was studied in \cite{shin2023photon}, and the simplified formula for the heralding efficiency $HE$ was given as
\begin{equation}\label{eq:HE}
   HE \approx \frac{(\alpha L)^2}{2(e^{\alpha L}-\alpha L - 1)}\,, 
\end{equation}
with $\alpha$ being an absorption coefficient. In the \textit{Results} section, we use this formula to compare the heralding efficiency for different waveguide structures and radii of curvature.

\subsection{Description of the considered waveguides}

In this subsection, we describe the different waveguides we consider in the following. To keep it rather general, we rely on a selection of waveguide examples, previously studied in the literature, made from different materials: silicon nitride buried in silicon dioxide \cite{levy2010cmos,moss2013new}, silicon photonic wires \cite{osgood2009engineering}, silicon on insulator (SOI) \cite{driscoll2011directionally,morrison2016four, Sakai2001}, and freeform waveguide made of IP-Dip photoresist (Nanoscribe GmbH) \cite{schmid2019optical}, which can be written with three-dimensional additive manufacturing techniques \cite{nesic2022transformation, schmidt2024}. Table \ref{tab:cases} lists the investigated waveguide systems. 
\begin{table}[!ht]
\caption{Investigated materials and structures}
\label{tab:cases}
\centering
\begin{tabular}{ c @{\hskip 0.5cm} c @{\hskip 0.5cm} c }
  core material & cladding material & core cross-section (w [\SI{}{\micro\meter}]$\times$ h [\SI{}{\micro\meter}])\\
 \hline
 %\multirow{4}{*}{free form} & 
    IP-Dip & $n_\text{clad}=1.36$ & 2 $\times$ 1.8\\  
   Si$_3$N$_4$ & SiO$_2$ & 2 $\times$ 1.8 \\
   Si$_3$N$_4$ & SiO$_2$ & 1.7 $\times$ 0.711 \\
   Si$_3$N$_4$ & SiO$_2$ & 1 $\times$ 1 \\
   Si$_3$N$_4$ & SiO$_2$ & 1 $\times$ 0.5 \\
   Si$_3$N$_4$ & SiO$_2$ & 0.46 $\times$ 0.3 \\
   Si & SiO$_2$ & 2 $\times$ 1.8\\
   Si & SiO$_2$ & 0.45 $\times$ 0.22\\
   Si & air & 1 $\times$ 0.32\\
 Si & air & 0.45 $\times$ 0.22\\
\end{tabular}
\end{table}
The considered waveguides differ not only in their core and cladding materials but also in their core sizes. To justify the choice of materials, we wish to stress that silicon is a popular core material due to its high nonlinearity and tight confinement of light due to high refractive index \cite{moss2013new}. Additionally, it has the advantage of combining electronics and photonics on the same chip \cite{moss2013new}. The major limitation of silicon is TPA. In response to this problem, silicon nitride platforms were developed, which reduce nonlinear losses at telecommunication wavelength \cite{morrison2016four,moss2013new}. Along with the high nonlinear materials, we also show one example of a photonic wire bond made from a polymer characterized by a weak nonlinearity and a rather low refractive index. That combination of material properties facilitates the discussion of the trade-off between loss and the possible enhancement of the nonlinear effects.

\section{Numerical details}\label{numerical_details}

In this section, we describe the computation steps. For each structure from Table \ref{tab:cases}, we start by computing the supported waveguide modes with the use of Ansys Lumerical \cite{lumerical} for wavelengths ranging from \SI{1510}{nm} to \SI{1600}{nm}, with \SI{5}{nm} step, which encompasses the common telecommunication wavelength \SI{1550}{nm}. The limits of the range are caused by difficulty in finding propagating modes for some materials and cross-sections for wavelengths outside the given range, especially for smaller bending radii. Each set of modes from the spectral range is computed separately for radii of curvature ranging from \SI{4}{\micro\meter} to \SI{100}{\micro\meter}. The exception is IP-Dip, for which the software could not find numerical modes for radii of curvature smaller than \SI{10}{\micro\meter} with the given settings. To numerically calculate the modes, the computational domain was limited to the waveguide core surrounded by the cladding of dimensions three times bigger than the core width. The FDE (finite difference eigenmode) simulation region was limited to 1.2 times the cladding width. To account for losses in bent waveguides, the PML (perfectly matched layers) were set on the boundaries of the computational domain $\mathcal D$. As the result of mode computation, the TE-polarized electric field of the fundamental mode, effective refractive index, effective area, loss, and group velocity of the mode are stored for further calculations.\\

In the case of silicon, silica, and silicon nitride, the refractive index values for particular frequencies are calculated from a Sellmeier equation, with the parameters determined by fitting the sample data points from the Ansys Lumerical database. To perform the analogous fitting for IP-Dip, the sample values of refractive index were taken from \cite{schmid2019optical}. The air and the cladding of the IP-Dip waveguide were assumed to be non-dispersive. The values and formulas for the components of the third-order susceptibility tensor were taken from \cite{boyd2008nonlinear}. The local phase and group velocities were calculated with formulas (68) and (69) from \cite{review_article}.\\

The nonlinear coefficients were computed with the equation \eqref{eq:gamma_simplified} for each combination of pump, signal, and idler wavelengths from the precomputed modes with the use of the formula for the effective area \eqref{eq:Aeff}. The maximum $\gamma_\text{SFWM}$, obtained for $\omega_\text{S}=\omega_\text{I}=\omega_\text{P}\approx193.42\cdot2\pi$ [THz], corresponding to wavelength $\lambda=1.55$ \SI{}{\micro\meter}, was chosen for comparison of different waveguides. The nonlinear coefficient was used to compute $M(\omega_\text{S},\omega_\text{I})$ from equation \eqref{eq:M} similarly for each combination of pump, signal, and idler wavelengths from the precomputed modes set.\\

The photon pair generation rate was computed from equation \eqref{eq:generation_rate} for every pump wavelength from the considered range. Due to the lack of data for all the frequencies (available just a limited set of discrete wavelengths), we modeled the Delta distribution with a Gaussian function $\delta(\Omega) = 1/\sqrt{\pi a^2}\exp(-\Omega^2/a^2)$, where the parameter $a$ was found empirically ($a=10^{12}$ Hz). The photon pair generation rate was calculated for the pump power \SI{100}{mW} injected into a waveguide of length \SI{15.7}{\micro\meter}, which corresponds to $90^\circ$ arc of radius \SI{10}{\micro\meter}. For the purpose of the photon pair generation rate comparison between the waveguides, the pump is fixed at \SI{1550}{nm}, and signal and idler are produced across the available spectral domain.

\section{Results}
In this section, we present the results of calculations for the examples listed in Table \ref{tab:cases}. We start with depicting the quantities used to calculate photon pair generation rate and their dependency on signal and idler wavelengths. Later, we discuss in detail one of the most efficient examples that we have looked at, i.e., the silicon nitride waveguide with a cross-section of 1.7 $\times$ 0.711 $\mu \mathrm{m}^2$ and the least efficient example, IP-Dip waveguide with a cross-section of 2.0 $\times$ 1.8 $\mu \mathrm{m}^2$, to show additional effects difficult to observe in almost lossless structures. An analysis with the discussion and physical interpretation follows, and finally, we summarize the results in Table \ref{tab:results}.\\
 
\begin{SCfigure}
		\includegraphics[width=8.6 cm,%0.47\textwidth,
        trim=30 155 30 155, clip]{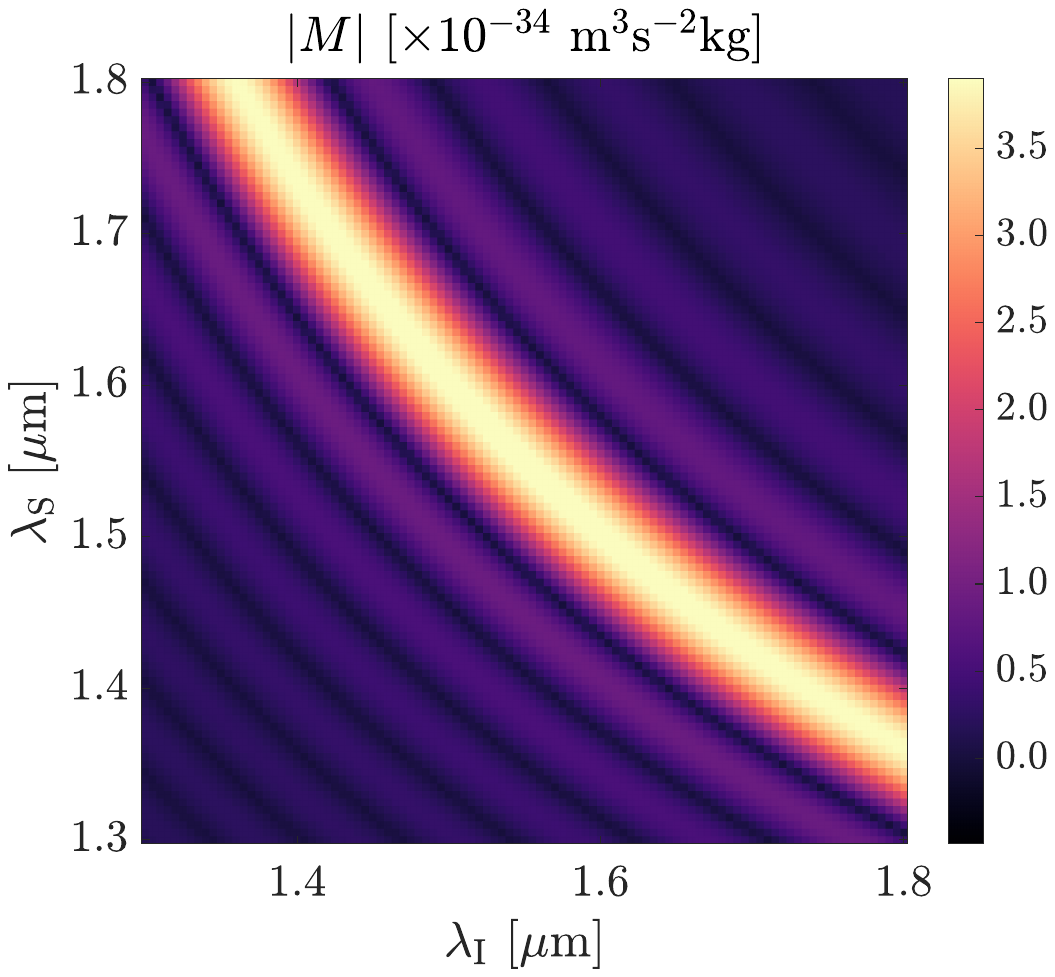}
	\captionsetup{width=\textwidth}
	\captionsetup{format=plain}
    \vspace{-0.2cm}
    \caption{The value of $|M|$ from equation (\ref{eq:M}) for a straight waveguide made of a silicon nitride core in a silica cladding for pump wavelength \SI{1.55}{\micro\meter} and different combinations of idler and signal wavelengths. Although the values in the image are always positive, the colorbar was shifted to increase the visibility of local maxima.}
	\label{fig:M_and_R}
\end{SCfigure}

Figure~\ref{fig:M_and_R} shows how the absolute value of the quantity $M$ depends on idler and signal wavelengths for one chosen pump wavelength. The local maxima occur for signal and idler wavelengths corresponding to modes that satisfy the phase-matching condition. For the plot in Fig.~\ref{fig:M_and_R}, the chosen wavelength span was \SI{1.3}{\micro\meter} -- \SI{1.8}{\micro\meter}. The values were calculated for 1.7 $\times$ 0.711 $\mu \mathrm{m}^2$ silicon nitride straight waveguides of length \SI{15.7}{\micro\meter}. The wavelength span for bent waveguides was chosen smaller as there may be a convergence problem for small radii and wavelength differing much from the central wavelength \SI{1.55}{\micro\meter}.\\

\begin{SCfigure}
    \centering
    \includegraphics[width=8.6 cm,%0.49\textwidth,
    trim=0 200 30 200, clip]{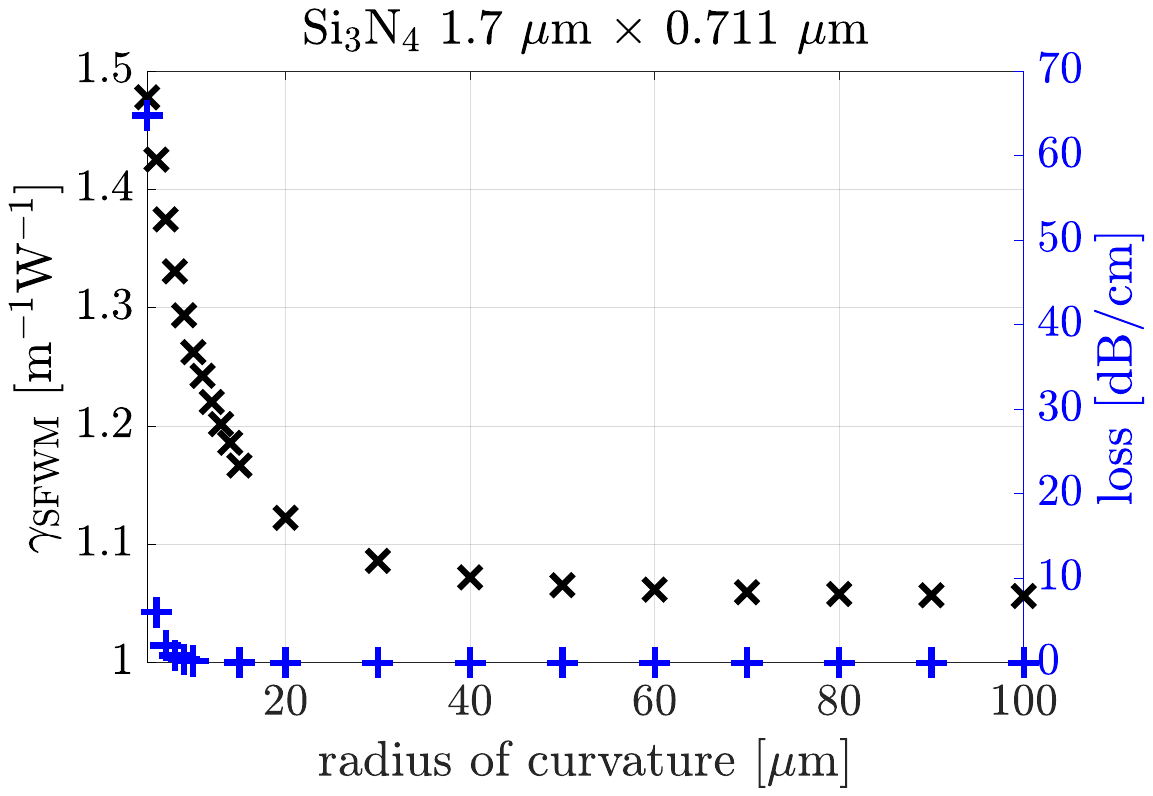}
    \vspace{-0.5cm}
    \caption{Nonlinear coefficient (black 'x') and loss (blue '+'), at pump wavelength \SI{1.55}{\micro\meter} as a function of the radius of curvature for silicon nitride 1.7$\times$\SI{0.711}{\micro\meter} core in silica cladding. Both the nonlinear coefficient and the loss increase with decreasing radius of curvature but at different rates.}
    \label{fig:gamma}
\end{SCfigure}

The first example of a bent waveguide we discuss in depth is silicon nitride in silica cladding, a very common material platform for photonic integrated circuits. In this case, the bending loss is minor, and the effective nonlinearity increases with increasing curvature. These dependencies are presented in Fig.~\ref{fig:gamma}. While both the nonlinear coefficient $\gamma_\text{SFWM}$ and bending loss increase as a function of the radius of curvature, the gain in nonlinearity is much higher and increases already for curvatures with minor bending losses.\\

\begin{figure}[ht]
	\centering
	\begin{subfigure}[c]{0.49\textwidth}
		\includegraphics[width=\textwidth,trim=0 200 40 200, clip]{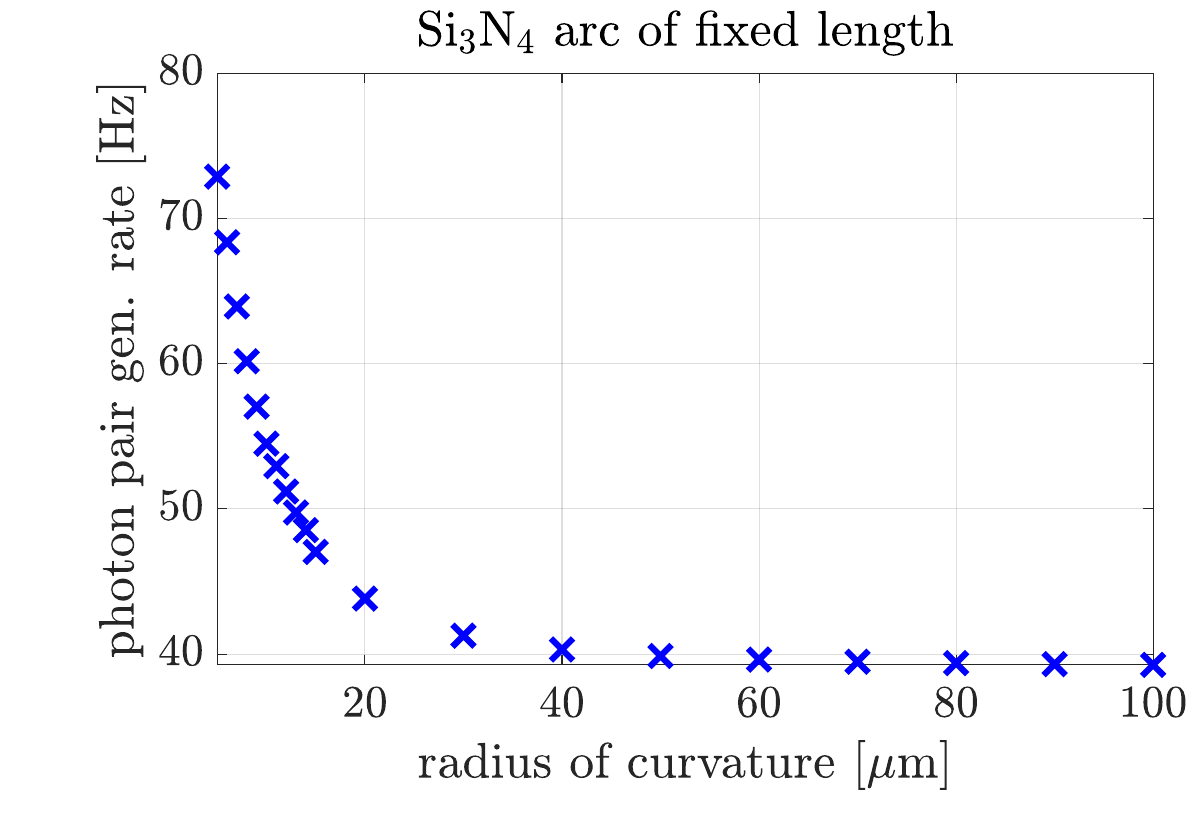}
    \caption{}
	\end{subfigure}
	\begin{subfigure}[c]{0.49\textwidth}
		\includegraphics[width=\textwidth,trim=20 205 20 200, clip]{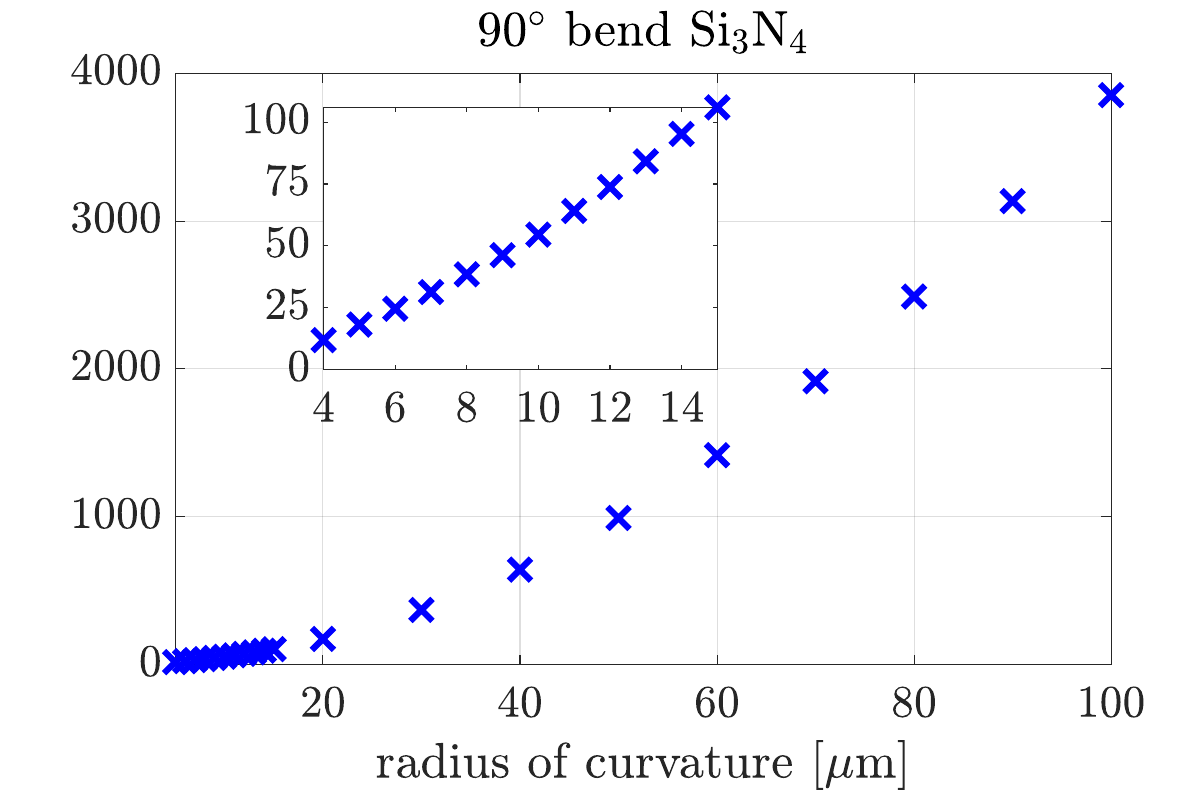}
    \caption{}
	\end{subfigure}
	\captionsetup{width=\textwidth}
	\captionsetup{format=plain}
    \caption{The photon pair generation rate as a function of radius of curvature for silicon nitride waveguide 1.7$\times$\SI{0.711}{\micro\meter} core in silicon dioxide cladding at pump wavelength \SI{1.55}{\micro\meter}. (a) The photon pair generation rate increases with increased bending for a fixed arc length, in this case, equal to \SI{15.7}{\micro\meter}. (b) The photon pair generation rate increases with increasing radius of curvature for 90$^\circ$ bow because the propagation length increases. The inset shows in detail the change of photon pair generation rate for radii of curvature between \SI{4}{\micro\meter} and \SI{15}{\micro\meter}.}
	\label{fig:peak_R_SOI}
\end{figure}

Consequently, the photon pair generation rate increases for waveguides that are stronger bent, as depicted in Fig.~\ref{fig:peak_R_SOI} (a), where the arc length is fixed and only the curvature changes. We see here clearly that the bending of the waveguide is beneficial. For the same length of the waveguide, the photon pair generation rate will be doubled in the case of \SI{4}{\micro\meter} bending radius as compared to the straight waveguide. From the different factors that affect the photon pair generation rate, we see that the effective nonlinearity increases with decreasing radius of curvature, which is obviously beneficial in improving the efficiency of the nonlinear processes. We explain this further below by the tighter localization of the mode in the curved waveguide. The increase in losses does not substantially degrade the photon pair generation rate. We stress that the bending loss would be more dominant for very small radii of curvature. Such a functional dependency can be better seen in waveguide examples where the modes are not that well localized, which we show in the following example of the IP-Dip waveguide.\\

Furthermore, we stress that the photon pair generation rate can be trivially enhanced by making the waveguides longer. This can be seen in Fig.~\ref{fig:peak_R_SOI} (b), where we show the photon pair generation rate for a fixed arc length of 90$^\circ$ bend as a function of the radius of curvature. Here, making the radius of curvature larger makes the waveguide longer, which allows us to generate a larger number of photon pairs. Nevertheless, in the light of a finite chip estate in a future generation of highly integrated photonic circuits, we should eventually ask ourselves which waveguide maximizes the observed effect for a given waveguide length. Comparing the two plots, we can conclude that the same generation rate reached by 90$^\circ$ bows of radii of curvature between \SI{10}{\micro\meter} and \SI{15}{\micro\meter} can be achieved with an \SI{15.7}{\micro\meter} long arc of smaller radius of curvature, below \SI{10}{\micro\meter}. This reduces the estate taken on the circuit board. In general, independent of the chosen length, superimposing the curvature as a design feature would always enhance the photon pair generation rate.\\

\begin{SCfigure}[][h]
   %th \centering
    \includegraphics[width=8.6 cm,%0.52\textwidth,
    trim=0 220 0 200, clip]{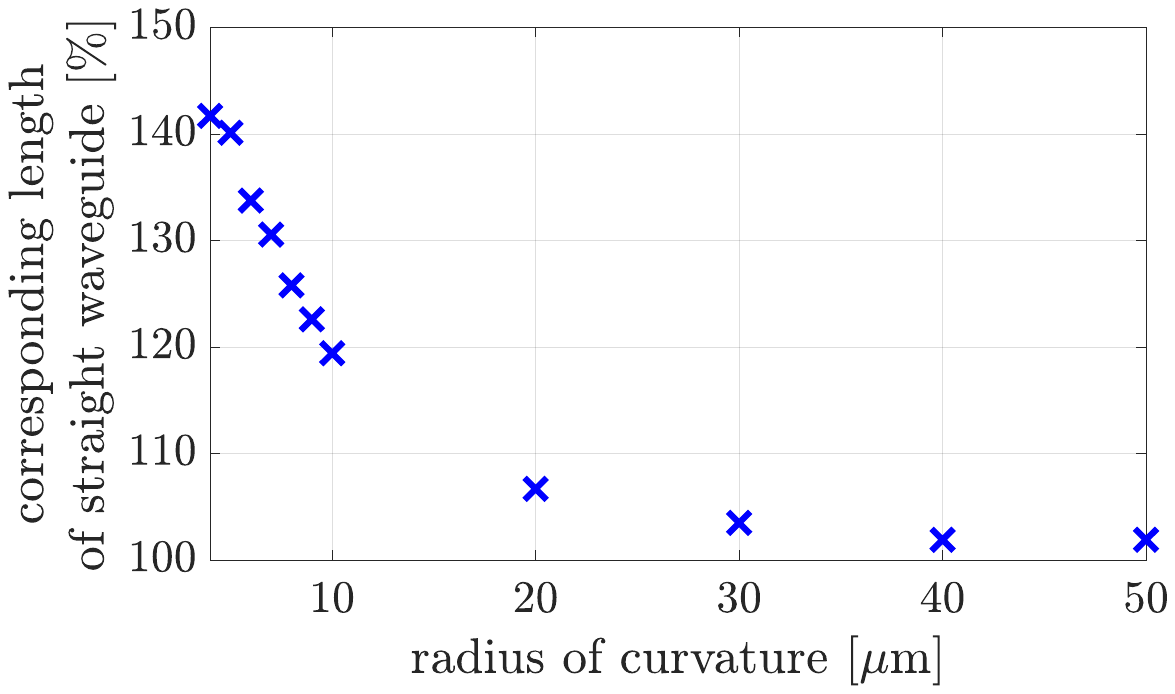}
    \caption{The fraction of the length of a straight waveguide to the corresponding length of a bent waveguide, resulting in the same photon pair generation rate. The values were calculated for silicon nitride bends of length \SI{15.7}{\micro\meter} and radii of curvature ranging from \SI{4}{\micro\meter} to \SI{50}{\micro\meter} and core cross-section \SI{1.7}{\micro\meter}$\times$\SI{0.711}{\micro\meter}.}
    \label{fig:equivalence}
\end{SCfigure}

While Fig.~\ref{fig:peak_R_SOI} shows that the photon pair generation rate increases both with decreasing radius of curvature and increasing length, Fig.~\ref{fig:equivalence} presents how much longer the straight waveguide should be to achieve the same generation rate as the bent one. It can be observed that for sharp bends, the corresponding straight waveguide should be almost 1.5 times longer. This fact can be used to increase the compactness of the waveguiding structures in the photonic integrated circuits. \\

\begin{SCfigure}[][h]
    \centering
    \includegraphics[width=0.5\textwidth,trim=0 200 30 200, clip]{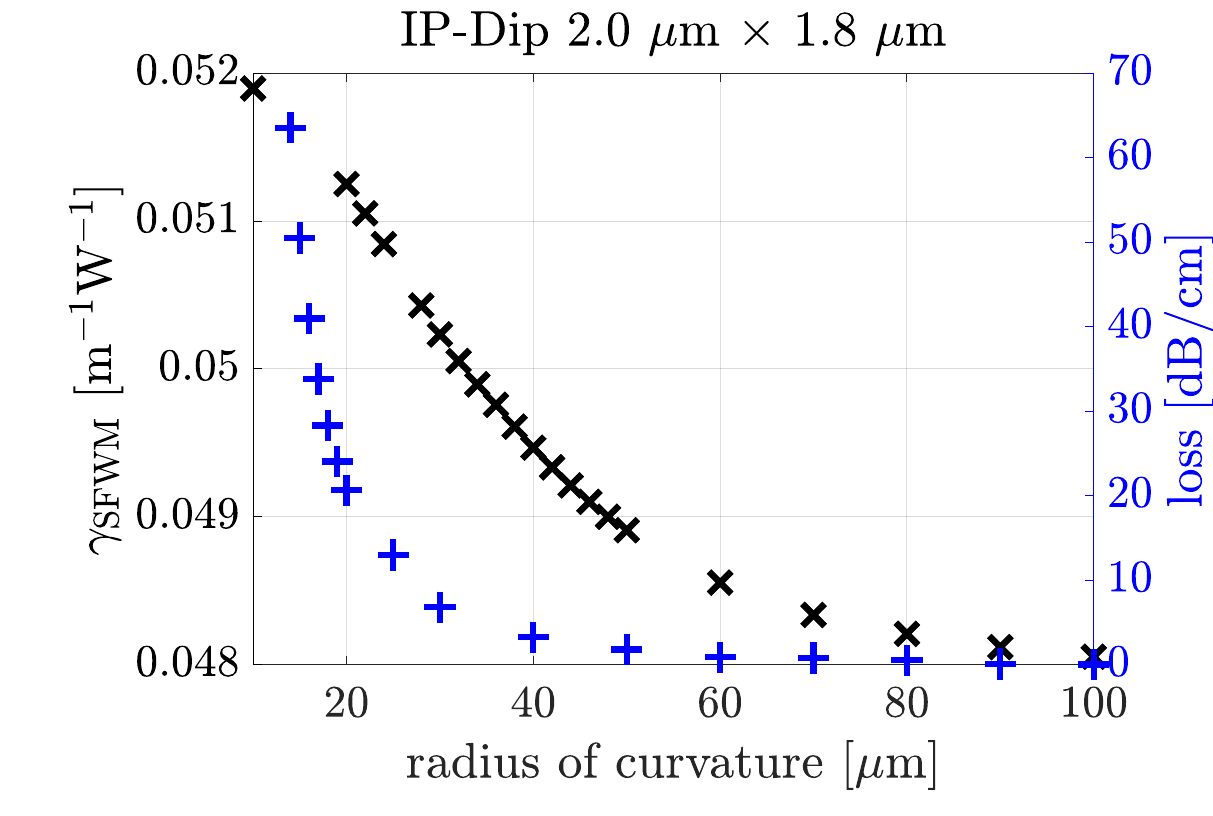}%0 200 30 200
    \caption{Nonlinear coefficient (black 'x') and loss (blue '+') as a function of the radius of curvature for IP-Dip \SI{2.0}{\micro\meter}$\times$\SI{1.8}{\micro\meter} core, at pump wavelength \SI{1.55}{\micro\meter}. Both the nonlinear coefficient and the loss increase with decreasing radius of curvature but at different rates.}
    \label{fig:gamma_IP_Dip}
\end{SCfigure}

\begin{SCfigure}[][h]
   %th \centering
    \includegraphics[width=0.5\textwidth,trim=0 205 30 200, clip]{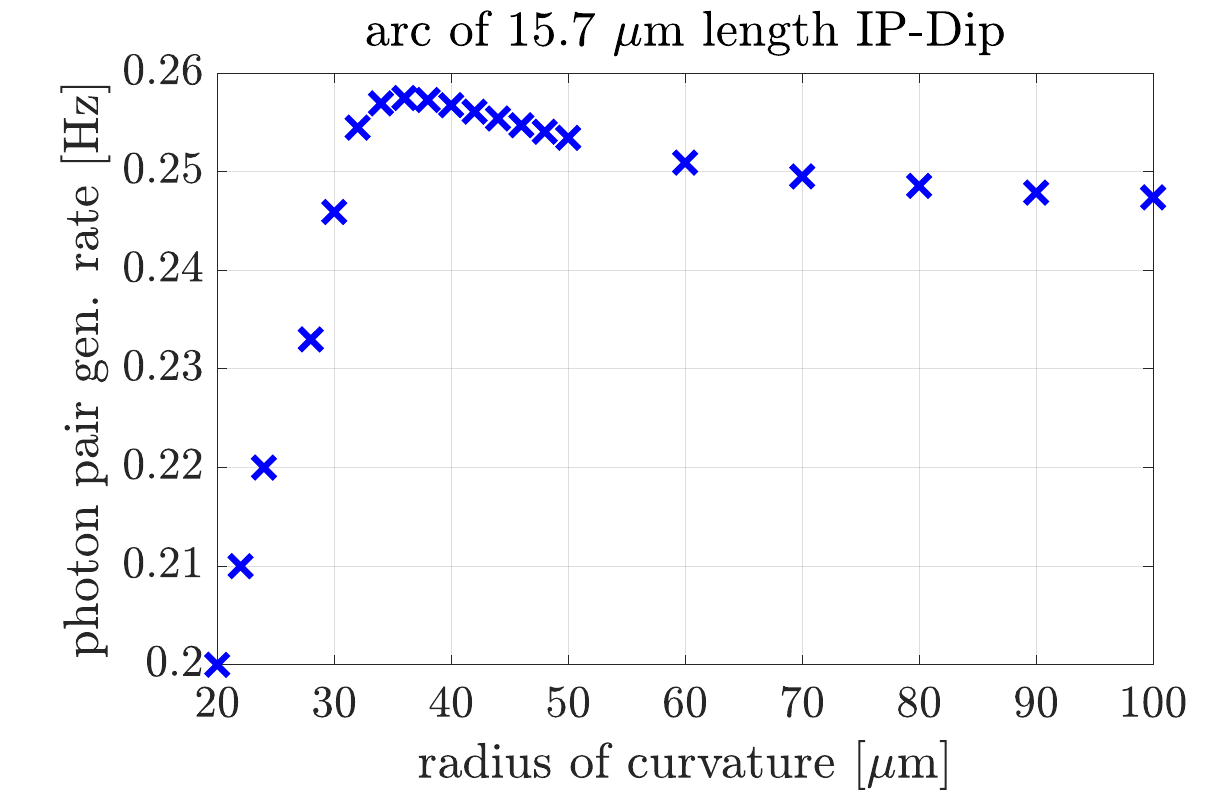}
    \caption{The photon pair generation rate at pump wavelength \SI{1.55}{\micro\meter} as a function of the radius of curvature for photonic wire bond fabricated with IP-Dip photoresist.}
    \label{fig:peak_R_IP_Dip}
\end{SCfigure}

A similar analysis can be performed for weakly nonlinear material IP-Dip. In Fig.~\ref{fig:gamma_IP_Dip}, one can see that the small nonlinear coefficient increases slightly with bending. At the same time, the losses start to be significant for a radius of curvature as big as \SI{40}{\micro\meter} and rise dramatically as the curvature increases. In the case of this core material, the loss influences the photon pair generation to a great extent, which is visible in Fig.~\ref{fig:peak_R_IP_Dip}.\\

\begin{SCfigure}[][h]
    \centering\includegraphics[width=0.5\textwidth,trim=0 220 0 200, clip]{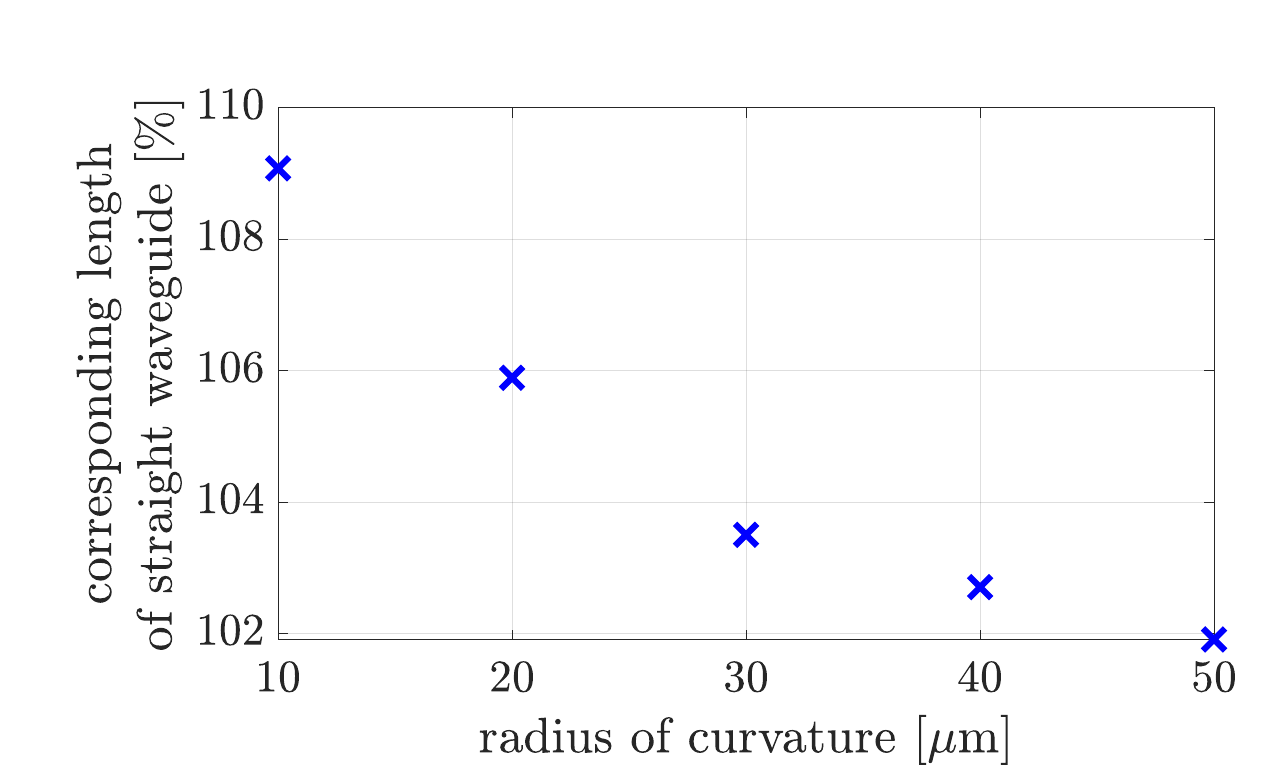}%0 200 30 200
    \caption{The fraction of the length of a straight waveguide to the corresponding length of a bent waveguide resulting in the same photon pair generation rate. The values were calculated for IP-Dip bends of length \SI{15.7}{\micro\meter} and radii of curvature ranging from \SI{10}{\micro\meter} to \SI{50}{\micro\meter} and core cross-section \SI{2.0}{\micro\meter}$\times$\SI{1.8}{\micro\meter}.}
    \label{fig:equivalence_IP_Dip}
\end{SCfigure}

Figure~\ref{fig:peak_R_IP_Dip} shows how the photon pairs generation rate changes for arcs of the same length \SI{15.7}{\micro\meter} as a function of different radii of curvature. Because the generation rate depends on propagation length, in this case, we study only the influence of bending on the result. The study shows that the generation rate increases with increasing radius up to \SI{36}{\micro\meter}. Then, a slight decay can be observed. It is clear that for small bending radii, the efficiency is impaired by loss. For higher radii of curvature, the generation rate converges to the limiting value for a straight waveguide because the nonlinear coefficient decreases, as shown already in Fig.~\ref{fig:gamma_IP_Dip}. Also, in the case of IP-Dip, it is possible to reduce the waveguide length by introducing bending, shown in Fig.~\ref{fig:equivalence_IP_Dip}. Due to induced losses, the length gain is very little and does not exceed 10~\%.\\

\begin{figure}[ht]
	\centering
	\begin{subfigure}[c]{0.49\textwidth}
		\includegraphics[width=\textwidth,trim=20 220 20 220, clip]{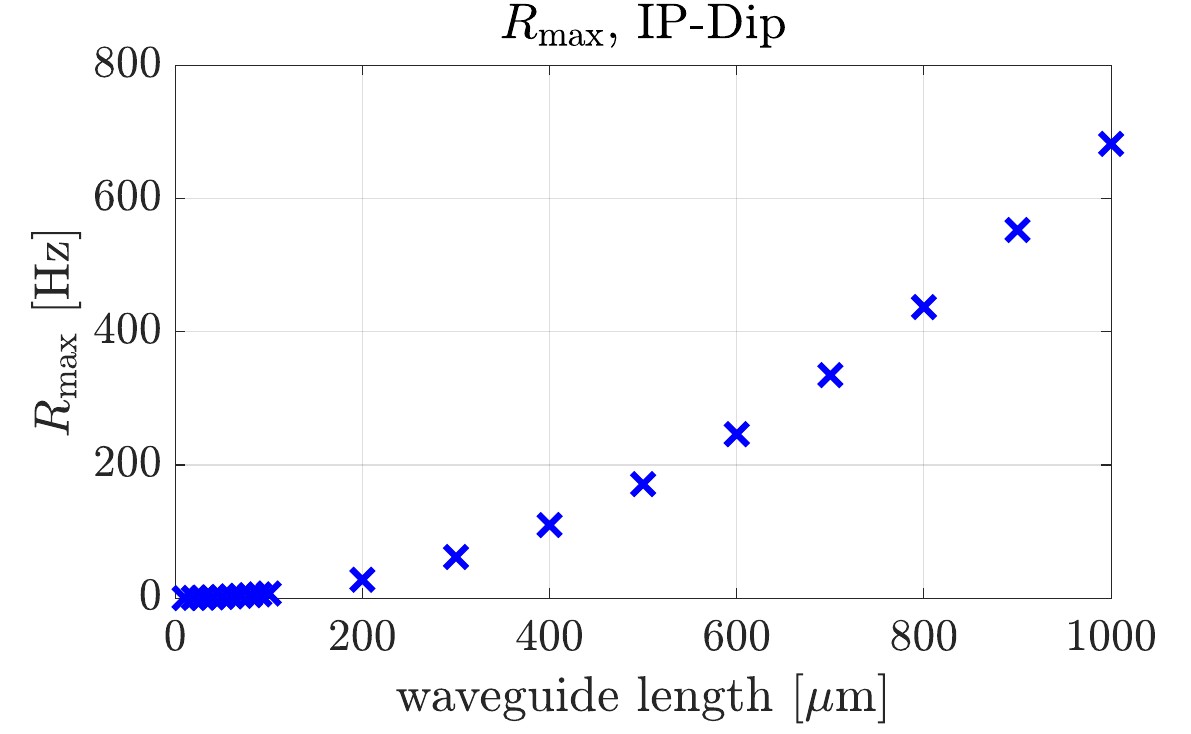}
        \caption{}
	\end{subfigure}
	\begin{subfigure}[c]{0.49\textwidth}
		\includegraphics[width=\textwidth,trim=20 220 20 220, clip]{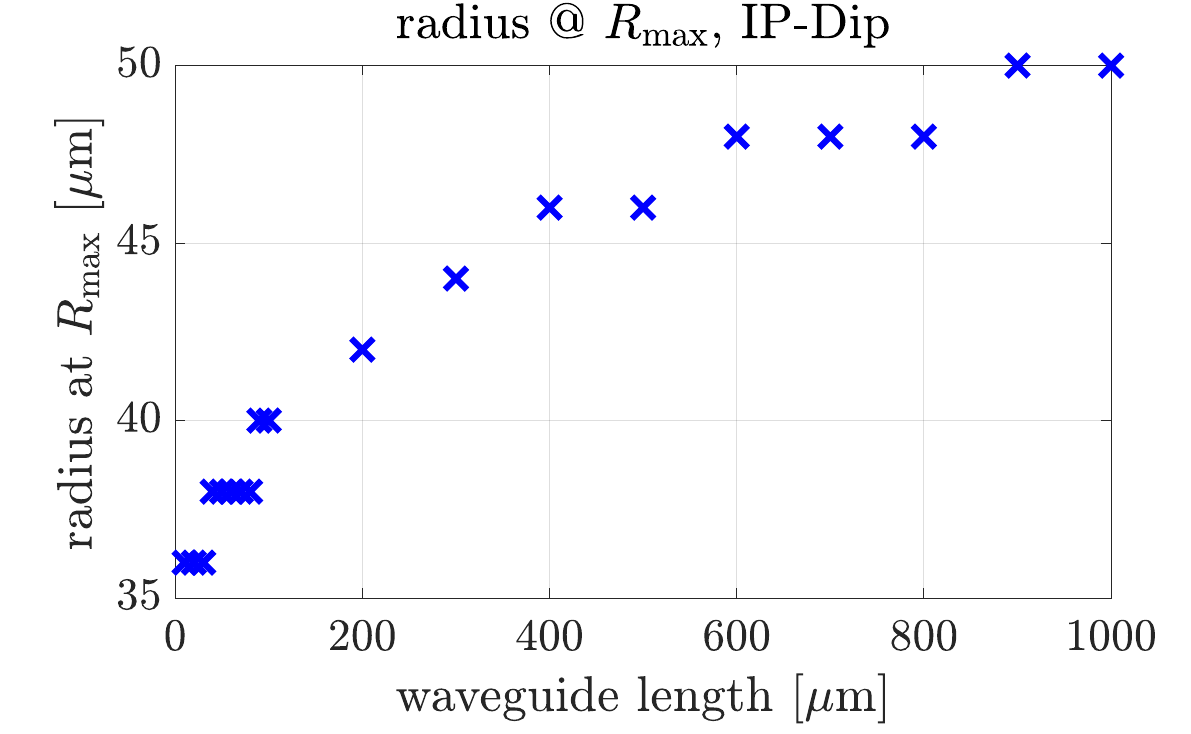}
	    \caption{}
    \end{subfigure}
	\captionsetup{width=\textwidth}    \captionsetup{format=plain}
    \caption{(a) Dependence of the generation rate on the waveguide length. (b) Radius for which the maximum generation rate is achieved as a function of waveguide length. Both plots were generated for IP-Dip.}\label{fig:dependence_on_length}
\end{figure}

It appears that the optimal radius of curvature may depend on the interaction length in the waveguide. Figure~\ref{fig:dependence_on_length} (a) shows the expected increase in photon pair generation rate as a function of waveguide length. In Fig.~\ref{fig:dependence_on_length} (b), one can see the corresponding radius of curvature for which the peak photon pair generation rate was achieved. The optimal radius of curvature increases with increasing interaction length. It is probably caused by the different rate at which losses and generation rate increase with the waveguide length.\\

%% TO-DO: show the table with the peak position for different materials and the value of peak generation rate

\begin{figure}[ht]
	\centering
    \vspace{-0.3cm}
	\begin{subfigure}[c]{0.7\textwidth}
		\includegraphics[width=\textwidth,trim=50 10 60 30, clip]{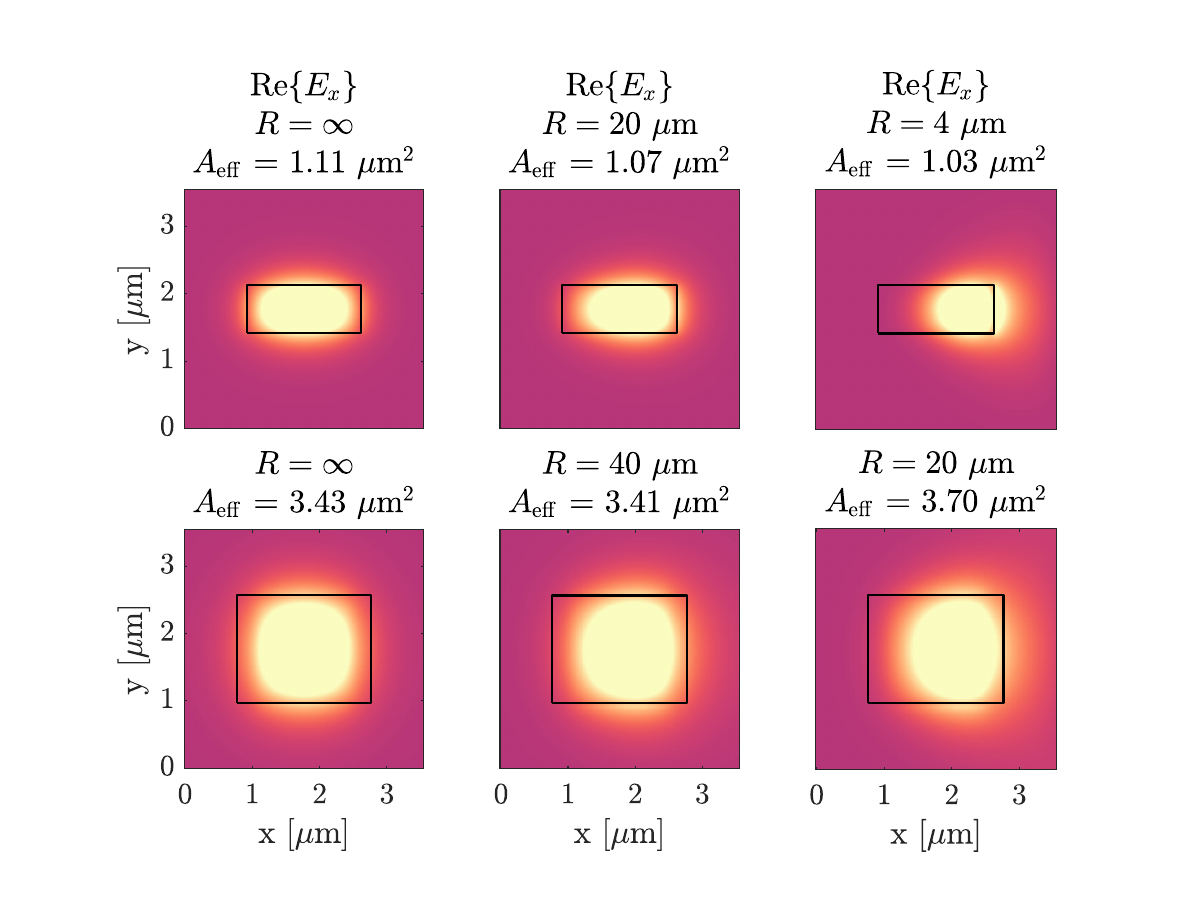}
	\end{subfigure}
	\hspace{-0.02\textwidth}
	\begin{subfigure}[c]{0.134\textwidth}%{0.0865\textwidth}
        %\vspace{3.5cm}
		\includegraphics[width=0.458\textwidth,trim=470 40 50 10, clip]{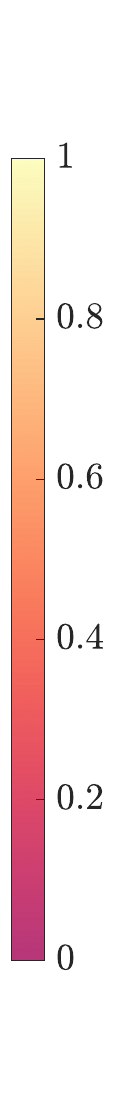}%trim=275 190 260 30
	\end{subfigure}
    \vspace{-0.3cm}
	\captionsetup{width=\textwidth}
	\captionsetup{format=plain}
    \caption{Mode profiles of the real part of the $x$-component of the normalized electric field obtained with Lumerical on the cross-section of waveguide for \SI{1.7}{\micro\meter}$\times$\SI{0.711}{\micro\meter} silicon nitride (top) and \SI{2.0}{\micro\meter}$\times$\SI{1.8}{\micro\meter} IP-Dip (bottom) for different radii of curvature, waveguide bent along $x$-axis. All fields are normalized to the maximum amplitude of the fundamental mode of the straight waveguide for the particular material. The core of the waveguide is marked with a black rectangle in scale. Although the computational domain depends on the core size, the cladding was cut to the same size for plotting. The effective area of the mode decreases with bending for silicon nitride. In the case of IP-Dip, it achieves its minimum for radius of curvature around \SI{40}{\micro\meter} and then increases due to leakage of the mode from the core to the cladding.}
	\label{fig:modes}
\end{figure}

In the following, we discuss the possible reasons for the enhancement of nonlinearity. Figure~\ref{fig:modes} shows mode profiles of the real part of the $x$-component of the normalized electric field on the waveguide cross-section and the corresponding effective areas of the modes. The upper row presents the cross-section of  \SI{1.7}{\micro\meter}$\times$\SI{0.711}{\micro\meter} silicon nitride, and the lower row contains the corresponding values for \SI{2.0}{\micro\meter}$\times$\SI{1.8}{\micro\meter} IP-Dip. The profiles in different columns are plotted for different bending radii. In the case of silicon nitride, increasing the curvature enhances the confinement of the electromagnetic field in the core and so reduces the effective area of the mode guided in the waveguide. In the case of IP-Dip, the effective area initially decreases, reaches the minimum, and then increases for very sharp bends. Decreasing the effective area causes an increase of the nonlinear coefficient $\gamma_\text{SFWM}$, as seen in equation \eqref{eq:gamma_simplified}.\\

\begin{figure}[ht]
	\centering
	\begin{subfigure}[c]{0.49\textwidth}
		\includegraphics[width=\textwidth,trim=20 220 20 220, clip]{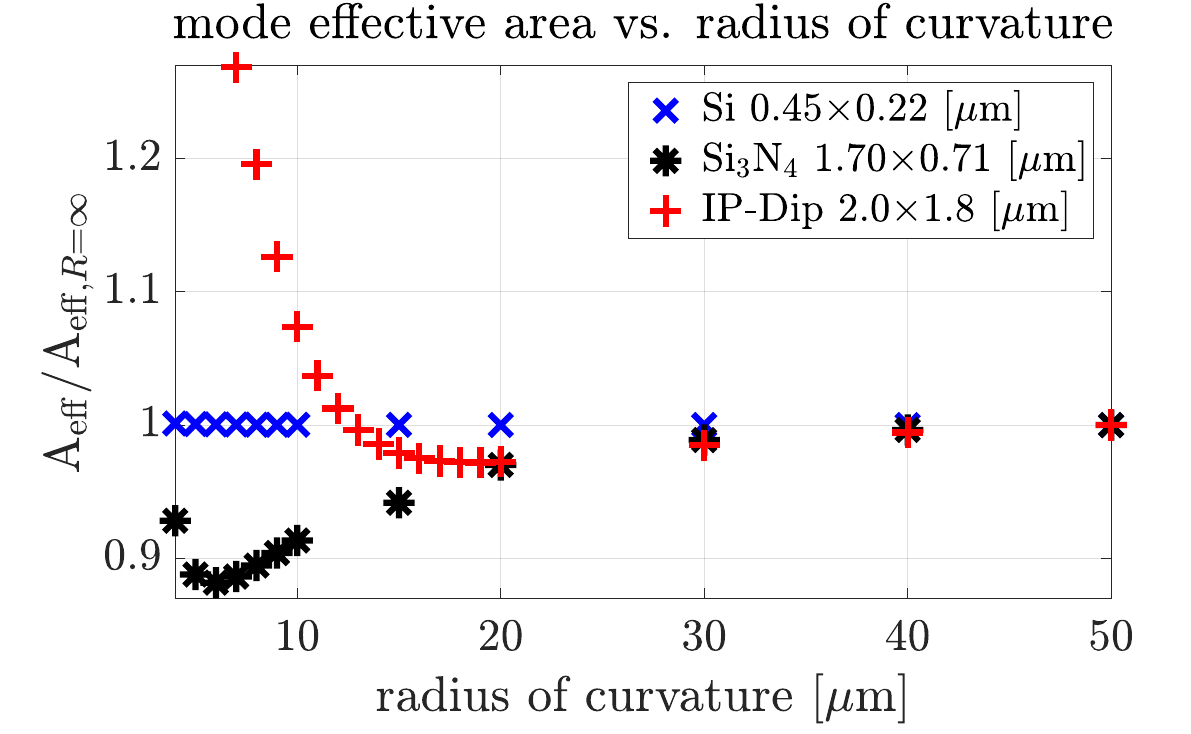}
        \caption{}
	\end{subfigure}
	\begin{subfigure}[c]{0.49\textwidth}
		\includegraphics[width=\textwidth,trim=20 220 20 220, clip]{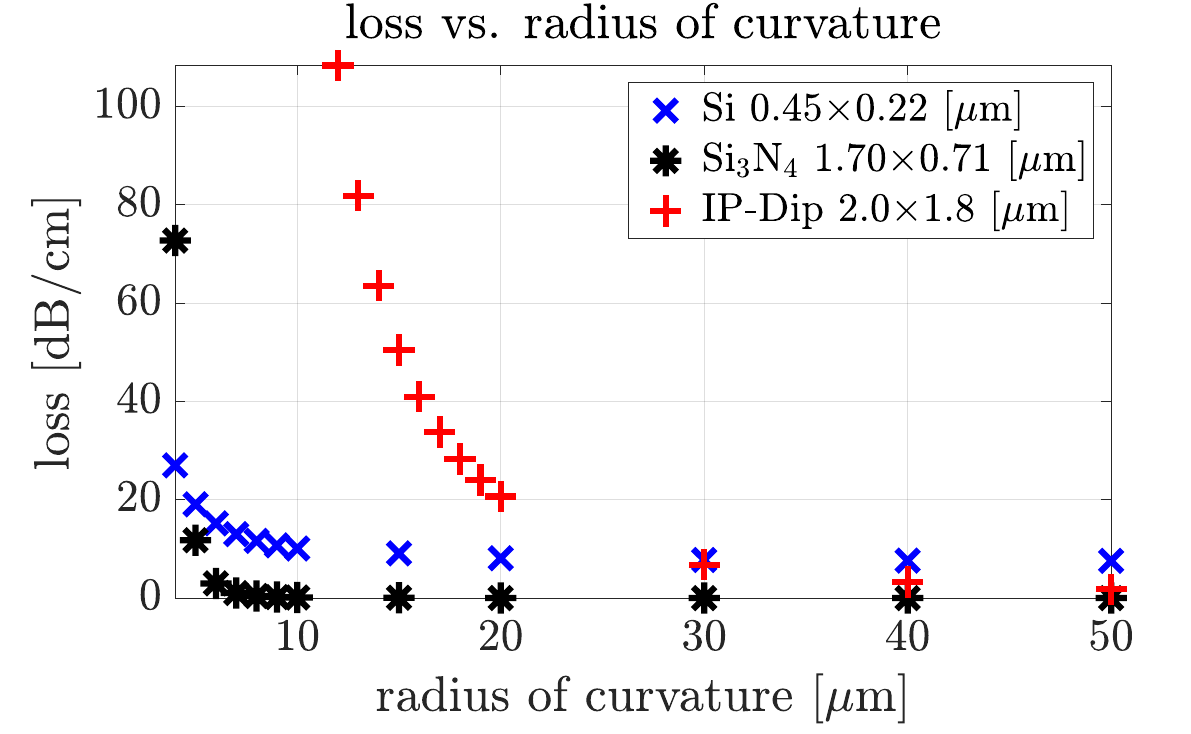}
        \caption{}
	\end{subfigure}
	\captionsetup{width=\textwidth}    \captionsetup{format=plain}
    \caption{(a) Effective area of the mode as a function of radius of curvature. (b) Dependence of the loss on the radius of curvature. The values were computed with Lumerical. The legend shows the materials and core cross-sections picked for the comparison. The values of mode effective area are normalized to the corresponding values of a straight waveguide.}\label{fig:loss_and_Aeff}
\end{figure}

Figure~\ref{fig:modes} suggests that the effective mode area changes with the bending radius of the waveguide. The explicit dependence of the area on the radius of curvature is shown in Fig.~\ref{fig:loss_and_Aeff} (a) for silicon nitride, IP-Dip, and also silicon for completeness, as it is the third material we study. One can see that for both silicon nitride and IP-Dip, it is possible to indicate the radius of curvature, for which the effective area of the guided mode is the smallest. In the case of silicon, the effective area of the mode is almost independent of the radius of curvature. The change of the effective area is compared to the losses in the waveguide as a function of the curvature, shown in Fig.~\ref{fig:loss_and_Aeff} (b). In the case of IP-Dip, the effective area decreases for radii of curvature smaller than \SI{50}{\micro\meter} until the minimum $A_\text{eff}$ is achieved for radii of curvature around \SI{20}{\micro\meter}. The mode radiates into the cladding for higher curvature, increasing the effective area. This effect is visible in Fig.~\ref{fig:loss_and_Aeff} (b), where the loss starts to increase strongly for the radii of curvature smaller than \SI{30}{\micro\meter}. Therefore, the optimal radius of curvature is to be expected somewhere in the range between \SI{30}{\micro\meter} and \SI{50}{\micro\meter}, as already shown in Fig.~\ref{fig:dependence_on_length} (b). A similar situation occurs for silicon nitride, but the mode characterized by a minimal effective area occurs at smaller radii, and the loss is much lower. Silicon behaves differently - the loss increases while there is no significant gain in the effective area of the mode. It results in regression of the photon pair generation rate, also visible in Table \ref{tab:results}, where the results for nonlinear coefficients and generation rate are collected for all the structures studied.\\

In Table \ref{tab:results}, the increment in photon pair generation rate can vary between 5~\% for IP-Dip and 272~\% for silicon. This property depends not only on the core and cladding material but also on the cross-section size. The bigger silicon nitride cores result in higher photon pair generation rate enhancement. In contrast, for very small cores, the losses outbalance the mode confinement, and the generation rate decreases with curvature. The enhancement effect cannot be observed due to the too low confinement of the mode to the core. In this case, the effective area of the mode tends to increase with decreasing radius of curvature, decreasing the nonlinear coefficient $\gamma_\text{SFWM}$. Similar dependency can be observed in the case of silicon and silicon on insulator.\\

\begin{table}[!ht]
\caption{Results of calculation for chosen waveguides of length \SI{15.7}{\micro\meter} and pump wavelength \SI{1.55}{\micro\meter}.}
\label{tab:results}
\centering
\begin{tabular}{ c | c @{\hskip 0.3cm} c@{\hskip 0.3cm} c @{\hskip 0.3cm}c @{\hskip 0.3cm}c}
 Waveguide type$^1$ & $\gamma_{\text{SFWM}, min}$$^2$ & $\gamma_{\text{SFWM}, max}$$^2$ & generation rate$^3$ [Hz] &increment$^4$ [\%] & peak radius$^5$\\
 \hline
 IP-Dip 2 $\times$ 1.8 & 0.048 & 0.052 & 0.25 & 5& 36\\  
 Si$_3$N$_4$ 2 $\times$ 1.8 & 0.33 & 0.53 & $4$ & 250 & 7\\
 Si$_3$N$_4$ 1.7 $\times$ 0.711 & 1.06 & 1.51 & $40$ & 188 & 4\\
 Si$_3$N$_4$ 1 $\times$ 1 & 0.79 & 0.95 & $21$ & 45 & 5\\
 Si$_3$N$_4$ 1 $\times$ 0.5 & -- & 1.64 & $90$ & -- & $\infty$\\
  Si$_3$N$_4$ 0.46 $\times$ 0.3 & -- & 0.33 & $6$ & -- & $\infty$\\
 Si 2 $\times$ 1.8 & 6.83 & 18.57 & $10^3$ & 272 & 4\\
 Si 0.45 $\times$ 0.22 & -- & 116.80 & $10^5$ & -- & $\infty$\\
 SOI 1 $\times$ 0.32 & 52.57 & 62.38 & $10^4$ & 40 & 4\\
 SOI 0.45 $\times$ 0.22 & 142 & 142 & $10^5$ & < 0.02 & 4\\ % gamma could be as big as 300 1/Wm [moss]
  \hline
  \multicolumn{4}{l}{$^1$ described by core material followed by cross-section size in \SI{}{\micro\meter}$\times$\SI{}{\micro\meter}.}\\
  \multicolumn{4}{l}{$^2$ values given in (Wm)$^{-1}$.}\\
  \multicolumn{4}{l}{$^3$ for straight waveguide of length \SI{15.7}{\micro\meter} and pump power \SI{100}{mW}.}\\
  \multicolumn{4}{l}{$^4$ increment of the generation rate with respect to a straight waveguide.}\\
  \multicolumn{5}{l}{$^5$ radius of curvature in \SI{}{\micro\meter} for which the maximum generation rate was achieved numerically.}
\end{tabular}
\end{table}

\begin{SCfigure}[][h]
    \centering\includegraphics[width=0.52\textwidth,trim=0 205 0 190, clip]{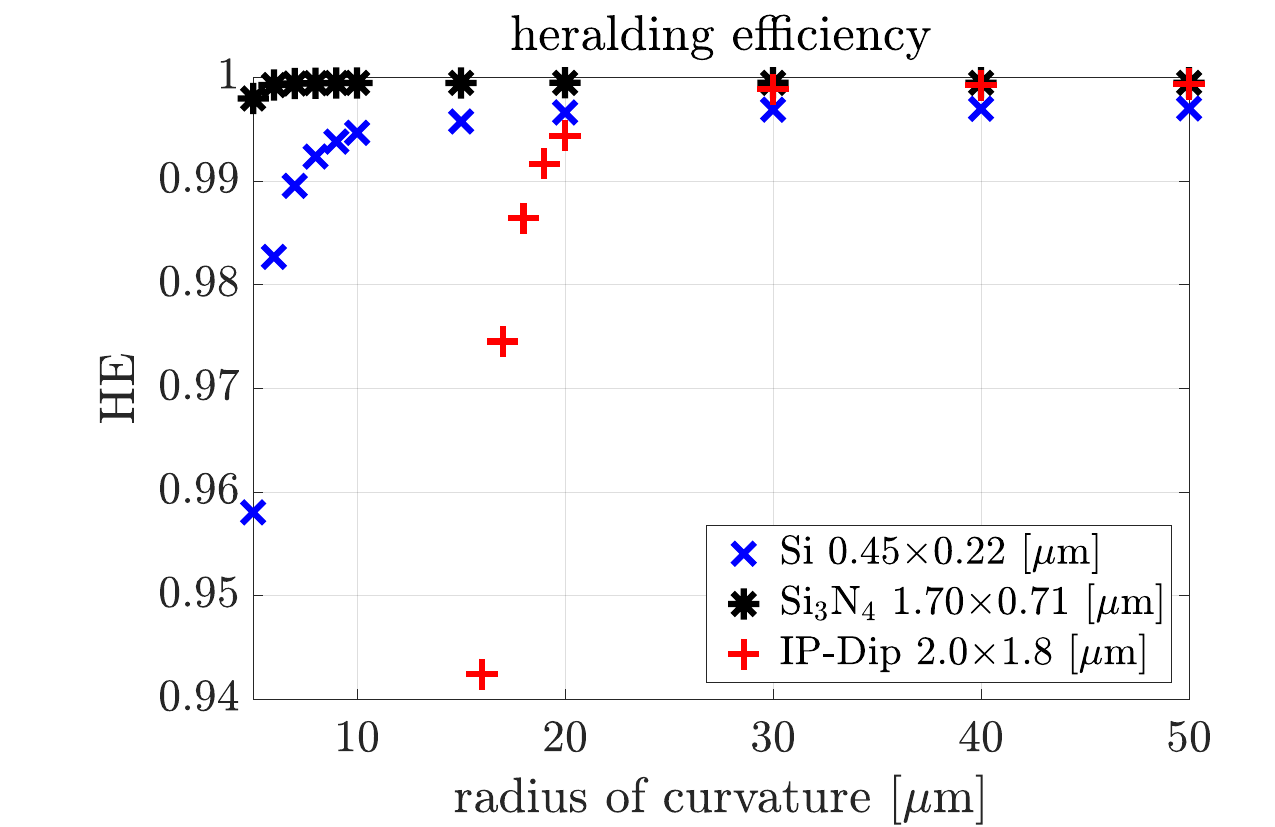}%0 200 30 200
    \caption{Heralding efficiency for different materials and cross-sections for interaction length \SI{15.7}{\micro\meter} as a function of radii of curvature between \SI{5}{\micro\meter} and \SI{50}{\micro\meter}.}\label{fig:HE}
\end{SCfigure}

The last feature we studied is the heralding efficiency, expressed with the formula \eqref{eq:HE}. We performed the calculations for exemplary silicon, silicon nitride, and IP-Dip waveguides, taking the imaginary part of the propagation constant as $\alpha$. The results are presented in Fig.~\ref{fig:HE}. For the sake of readability, the values of radii of curvature were limited to \SI{5}{\micro\meter} as the heralding efficiency dropped below 0.8 for higher curvatures. The $HE$ of silicon starts at 0.958 for a radius of curvature \SI{5}{\micro\meter} and 0.998 for silicon nitride. The photoresist IP-Dip exceeds the heralding efficiency of 0.98 for bending radii greater than \SI{18}{\micro\meter}. Depending on the threshold heralding efficiency in the experiment, the proper radius of curvature can be chosen for the waveguide design.

\section{Conclusions}
The enhancement of the effective nonlinearity in waveguides by exploiting their curvature has been discussed at a qualitative and quantitative level. We investigated how the photon pair generation rate in a spontaneous four-wave mixing process changes with an increase in the curvature of the waveguides. The process is technologically important, as it is used to generate quantum states of light that are a source for a future generation of on-chip quantum technological devices. We considered third-order strongly nonlinear core materials like silicon nitride and silicon and compared their behavior to weakly nonlinear material IP-Dip. We showed that increasing the curvature results in stronger mode confinement in the waveguide core, reducing the effective area of the mode and increasing the nonlinear coefficient. At the same time, the scattering losses increase but at a rate different from the nonlinear coefficient, which allows an indication of the optimal radius of curvature for the photon pair generation rate. The optimal curvature may depend on the interaction length, which should be considered for the design. Also, the required heralding efficiency should be taken into account for experiments. \\

The bending of the waveguide has a clear advantage for the confinement of guided modes. The spatial confinement of light enhances the nonlinear interaction strength. This effect has the potential to reduce the interaction lengths in the waveguides. We showed the photon pair generation enhancement for very short waveguides. In practical cases, where much longer waveguides are fabricated, the space gain may be much higher, assuming the low-loss range of the radii of curvature.

\section{Acknowledgements}
The first author (M.P.) highly appreciates the supervision of Prof. John E. Sipe during the research stay at the University of Toronto and is thankful to him for the numerous meetings and discussions during the work on this project. M.P. is thankful to the Institute of Photonics and Quantum Electronics at Karlsruhe Institute of Technology for sharing the Ansys Lumerical license. M.P. is also thankful to Colin Vendromin from the University of Toronto for useful exemplary scripts in Ansys Lumerical and helpful discussions. The authors are grateful to Dr. Frank Setzpfandt from Friedrich Schiller University Jena for the insightful discussions and hints regarding the experimental aspects of this work. The research stay of M.P. at the University of Toronto was funded by the Mitacs Globalink Research Award and supported by the Karlsruhe House of Young Scientists (KHYS). M.P. acknowledges support by the NHR@KIT program. This work and part of the research stay in Toronto was funded by the Deutsche Forschungsgemeinschaft (DFG, German Research Foundation) Project-ID 258734477 -- SFB -- 1173.

\end{CJK*}
\bibliographystyle{ieeetr}
\bibliography{bibliography}
\end{document}